\documentclass[preprints,article,accept,moreauthors,pdftex]{Definitions/mdpi}
\usepackage{mathtools}
\makeatletter
\newcommand{\vast}{\bBigg@{4}}
\newcommand{\Vast}{\bBigg@{5}}

\makeatother

\firstpage{1} 
\pubvolume{xx}
\issuenum{1}
\articlenumber{5}
\pubyear{2020}
\copyrightyear{2020}
\history{Received: date; Accepted: date; Published: date}
\Title{Modeling atom interferometry experiments with Bose-Einstein condensates in power-law potentials}

\Author{
S. Thomas$^1$, C. Sapp$^1$, C. Henry$^1$, A. Smith$^1$, C.A. Sackett$^2$, C.W. Clark$^3$ and M. Edwards$^{1\ast}$}

\AuthorNames{Stephen Thomas, Colson Sapp, Charles Henry, Andrew Smith, Charles A. Sackett, Charles W.\ Clark, and Mark Edwards}
\address{\quad $^1$Department of Physics and Astronomy, Georgia Southern University, Statesboro, GA 30458\\
\quad $^2$Department of Physics, University of Virginia, Charlottesville, VA 22904\\
\quad $^3$Joint Quantum Institute, National Institute of Standards and Technology and University of Maryland, College Park, MD 20742}

\corres{Correspondence: edwards@georgiasouthern.edu}

\abstract{Recent atom interferometry (AI) experiments involving Bose--Einstein condensates (BECs) have been conducted under extreme conditions of volume and interrogation time.  Numerical solution of the standard mean-field theory applied to these experiments presents a nearly intractable challenge. We present an approximate variational model that provides rapid approximate solutions of the rotating-frame Gross--Pitaevskii equation for a power-law potential. This model is well-suited to the design and analysis of AI experiments involving BECs that are split and later recombined to form an interference pattern.  We derive the equations of motion of the variational parameters for this model and illustrate how the model can be applied to the sequence of steps in a recent AI experiment where BECs were used to implement a dual-Sagnac atom interferometer rotation sensor.  We use this model to investigate the impact of finite-size and interaction effects on the single-Sagnac-interferometer phase shift.}


\begin{document}
%%%%%%%%%%%%%%%%%%%%%%%%%%%%%%%%%%%%%%%%%%

%%%%%%%%%%%%%%%%%%%%%%%%%%%%%%%%%%%%%%%%%%%%%%%%%%
%                  Introduction                  %
%%%%%%%%%%%%%%%%%%%%%%%%%%%%%%%%%%%%%%%%%%%%%%%%%%
\section{Introduction}
\label{intro}

Ultracold atom interferometry (AI) has become a mature technology over the past 25 years~\cite{Bongs2019}.  Today it has many applications including precision metrology~\cite{kibble_balance, Richard2018, big_G, Menoret2018, Weiss1993, Andreas2011, Bouchendira2011, Hanneke2008, Fixler2007}, quantum sensing~\cite{Bonnin2013, Dimopoulos2008a, Hogan2016, Hogan2009, Dickerson2013, Zhou2011, Barrett2016, Antoine2003, Geiger2011, Amaro-Seoane2012, Chaibi2016, PhysRevLett.124.120403}, and tests of fundamental physics~\cite{Niebauer1987, Touboul2017, Zhang2015, Fray2004, Tarallo2014, Schlippert2014, Rosi2017a, Rosi2017b, Geiger2018, Elder2016, Hamilton2015, Jaffe2017, Strigari2013, Arvanitaki2018, Hees2016, Overstreet2018}.  Precision metrology applications include measurements of the fine-structure and universal gravitation fundamental constants, geophysical measurements, and measuring the local acceleration of gravity as part of the new Kibble-balance kilogram standard.  AI quantum sensors measure  rotation and acceleration for precision navigation and serve as gravity gradiometers for geodesy and civil engineering. Atom interferometers have also been proposed for gravity-wave detection.  Probes of fundamental physics include dark-matter and dark-energy searches as well as tests of Einstein's Equivalence Principle.

Many AI experiments involve splitting and recombining a Bose--Einstein condensate (BEC). Early measurements demonstrated that BECs were capable of interference~\cite{Andrews637, PhysRevLett.85.2040, PhysRevLett.83.3112}. More recently, BEC atom interferometers have been used for measuring the fine-structure constant~\cite{PhysRevLett.89.140401} and as gravimeters~\cite{PhysRevA.84.033610}.  Condensates have been used in interferometers on atom chips~\cite{Abend2016}, have been confined in large ring potentials~\cite{Bell2016,2019Natur.570..205P, gerhard1,de_Go_r_de_Herve_2021}, and launched through waveguide painted-potentials~\cite{Ryu2015}.  %Finally, condensates have been created aboard the International Space Station in the Cold Atom Laboratory (CAL) facility deployed by the National Aeronautics and Space Administration (NASA)~\cite{nasa_cal}.  The CAL was recently upgraded to include hardware capable of performing AI operations.
Finally, extraterrestrial AI uses BEC as a source. In 2017, the MAIUS-1 sounding-rocket mission produced BEC and conducted AI experiments above the Kármán line, between 100 and 243 km above the Earth's surface.\cite{2018MAIUS,BECCAL}. In 2020, a BEC of some 50,000 $^{87}$Rb atoms was produced aboard the Earth-orbiting International Space Station.~\cite{Aveline1,Aveline2} There is an ongoing effort to implement BEC-enabled AI in that environment.~\cite{nasa_cal}

Using condensates in AI processes confer the advantages of narrow momentum distributions and better signals due to their higher density relative to above-$T_{c}$ gases. Thus using condensates in AIs hold high promise for accurate measurements~\cite{PhysRevA.84.043643}. High densities have the disadvantage that interactions can cause phase diffusion that tends to reduce the contrast of the interference pattern generated by overlapping BEC clouds~\cite{Grond_2010}.  Furthermore, mean-field repulsion of overlapping clouds can alter their relative velocity which can change the frequency of the interference pattern~\cite{bragg_proto}.  Indeed many BEC AI experiments use small condensates in order to minimize the effects of interactions. While there has been a good deal of theoretical work done on the effect of interactions in BEC AIs~\cite{PhysRevA.84.043643, olshanii_dunjko, PhysRevA.77.043604, PhysRevA.80.063617}, few have used accurate solutions of the Gross--Pitaevskii equation (GPE) in their analysis.

The time-dependent Gross--Pitaevskii equation is the standard mean-field theory that governs the behavior of confined ultracold gases at near-zero temperatures~\cite{Pitaevskii_and_Stringari}.  There is every reason to think that it will apply to BEC AI systems even in the most extreme cases where the condensate is split into multiple high-momentum clouds which are then allowed to separate to a distance that is orders of magnitude larger than the original condensate size and where the system evolution time is long. Precision solutions of the GPE should thus be able accurately to account for the effects of interactions.  The reason that previous analyses of BEC AIs have not solved the GPE numerically is that the extreme conditions (large-momentum, large-volume, long interrogation times) of these AIs are beyond the current state of the art of known algorithms for solving the GPE on a practical timescale~\cite{gpe_solver}.  

In this paper we analyze the operation of a recent dual-Sagnac atom interferometry rotation sensor experiment\,\cite{sackett_dual_sagnac} using a variational model approximation\,\cite{PhysRevA.99.043615} to the rotating-frame Gross--Pitaevskii equation.  We use this model to study interaction effects by increasing the number of atoms in the condensates used in the experiment.  We also study the effects of the presence of anharmonic terms in the external potential. In Section\,\ref{var_solver} we review the variational model and write down the equations of motion for the variational parameters.  In Section\,\ref{sackett_ai} we present the AI sequence carried out in the dual-Sagnac rotation speed measurement.  In Section\,\ref{power_law}, we write down the equations of motion for the variational parameters contained in trial wave condensate wave function.  In Section\,\ref{stopped_atom_fraction}, we derive exact expressions for the stopped-atom fraction in terms of the model variational parameters. This stopped-atom fraction is necessary for experimental extraction of the speed of the rotating frame from the data.  In Section\,\ref{approx_solutions} we derive approximate formulas for this quantity for zero-rotation speed and for a harmonic potential when cloud-cloud interactions are neglected. In Section\,\ref{sim_study} we present the results of a set of simulations designed to study the effects of interactions and anharmonic terms in the potential. Finally we summarize in Section\,\ref{summary}.

%%%%%%%%%%%%%%%%%%%%%%%%%%%%%%%%%%%%%%%%%%%%%%%%%%
%           Variational GPE Solver Model         %
%%%%%%%%%%%%%%%%%%%%%%%%%%%%%%%%%%%%%%%%%%%%%%%%%%
\section{Variational GPE Solver Model}
\label{var_solver}

We have developed and implemented~\cite{PhysRevA.99.043615} a variational model that provides rapid approximate solutions for the rotating-frame Gross--Pitaevskii equation (RFGPE) given by~\cite{Pitaevskii_and_Stringari}
\begin{eqnarray}
i\hbar\frac{\partial\Phi}{\partial t}
&=&
-\frac{\hbar^{2}}{2M}\nabla^{2}\Phi + 
V_{\rm ext}({\bf r},t)\Phi +
g_{3D}N\left|\Phi\right|^{2}\Phi +
i\hbar{\bf \Omega}\cdot\left({\bf r}\times{\bf\nabla}\right)\Phi,
\end{eqnarray}
where $\Phi({\bf r},t)$ is the condensate wave function, $M$ is the mass of a condensate atom, $N$ is the number of atoms in the condensate, $g_{3D}=4\pi\hbar^{2}a_{s}/M$ measures the strength of the atom-atom scattering with $a_{s}$ being the scattering length, $V_{\rm ext}({\bf r},t)$ is the potential exerted on a condensate atom by external fields, and ${\bf\Omega}$ is the angular velocity of the rotating frame. 

Our variational model is based on the Lagrangian Variational Method (LVM)~\cite{PhysRevLett.77.5320,PhysRevA.56.1424} which provides equations of motion for time-dependent variational parameters appearing in a trial wave function.  The equations of motion for these variational parameters are derived by integrating the Lagrangian density over all space yielding the ordinary Lagrangian and then using the standard Euler--Lagrange equations, to produce an equation of motion for each variational parameter.

In our model the trial wave function represents each of the $N_c$ wave packets as a 3D Gaussians. The mathematical form is the superposition:
\begin{equation}
\Psi({\bf r},t) = 
\frac{1}{\sqrt{N_{c}}}\sum_{j=1}^{N_{c}}
\textcolor{black}{A_{j}(t)}
e^{g_{j}({\bf r},t)}
\label{trial_wf_3d}
\end{equation}
where
\begin{eqnarray}
g_{j}({\bf r},t) 
&=&
\sum_{\eta = x,y,z}
\left(
-\frac
{(\eta-\textcolor{black}{\eta_{j}(t)})^{2}}
{2\textcolor{black}{w_{j\eta}(t)}^{2}} +
i\textcolor{black}{\epsilon_{j\eta}(t)}\eta + 
i\textcolor{black}{\beta_{j\eta}(t)}\eta^{2}
\right).
\label{trial_exp_3d}
\end{eqnarray}
The variational parameters are the center coordinates $\eta_j$, the widths $w_{j\eta}$, and linear $\epsilon_{j\eta}$ and quadratic $\beta_{j\eta}$ phase parameters for each cloud.  These are indicated above as explicit functions of time.  

Here we make two assumptions about the physical system which have material effects on the values of the variational parameters.  These are as follows:
\begin{enumerate}
  \item We assume that each of the $N_{c}$ clouds are moving at sufficiently
  different velocities such that any integral of a quantity containing a factor 
  like $\exp\left\{i\left(k_{j\eta}-k_{j^{\prime}\eta}\right)\eta\right\}$ 
  where $j\ne j^{\prime}$ can be neglected.  If the clouds move with sufficiently
  different velocities, these factors will be rapidly oscillating and their 
  integrals can be neglected.
  \item The number of atoms in each cloud is fixed.  Clouds do not lose or exchange
  atoms.  
\end{enumerate}
We can use these assumptions plus the normalization condition on the trial wave function to derive conditions that constrain the values of the $A_{j}$.  Our assumption that the number of atoms in each cloud is fixed adds the further restriction that each cloud is individually normalized.  The result is
\begin{equation}
\left|A_{j}(t)\right|^{2}
\pi^{3/2}
w_{jx}(t)
w_{jy}(t)
w_{jz}(t) = 1,
\quad
j = 1,\dots,N_{c}.
\label{cloud_norm}
\end{equation}

The equations of motion are a pair of second-order ordinary differential equations for the cloud centers and widths as well as expressions for the $\beta_{j\eta}$ and the $\epsilon_{j\eta}$ in terms of the centers, widths and their first derivatives:
\begin{subequations}
\begin{align}
\label{xj_eom_final}
\ddot{x}_{j} &=
2\Omega_{z}\dot{y}_{j} +
\Omega_{z}^{2}x_{j} -
\frac{1}{M}
\frac{\partial U^{(3D)}}{\partial x_{j}},\\
\label{yj_eom_final}
\ddot{y}_{j} &=
-2\Omega_{z}\dot{x}_{j} +
\Omega_{z}^{2}y_{j} -
\frac{1}{M}
\frac{\partial U^{(3D)}}{\partial y_{j}},\\
\label{zj_eom_final}
\ddot{z}_{j} &= 
-\frac{1}{M}
\frac{\partial U^{(3D)}}{\partial z_{j}},\\
\label{wjeta_eoms_final}
\ddot{w}_{j\eta} &=
\frac{\hbar^{2}}{M^{2}}
w_{j\eta}^{-3} - 
\frac{2}{M}
\frac{\partial U^{(3D)}}
{\partial w_{j\eta}},\\
\label{betajeta_eoms_final}
\beta_{j\eta} &=
\frac{M}{2\hbar}
\frac{\dot{w}_{j\eta}}{w_{j\eta}},\\
\label{alphajx_eom_final}
\epsilon_{jx} &=
\tfrac{M}{\hbar}(\dot{x}_{j} - \Omega_{z}y_{j}) - 
2\beta_{jx}x_{j},\\
\label{alphajy_eom_final}
\epsilon_{jy} &=
\tfrac{M}{\hbar}(\dot{y}_{j} + \Omega_{z}x_{j}) - 
2\beta_{jy}y_{j},\\
\label{alphajz_eom_final}
\epsilon_{jz} &=
\tfrac{M}{\hbar}\dot{z}_{j} - 
2\beta_{jx}x_{j},\\
\eta &= x,y,z\quad j=1,\dots,N_{c}\nonumber
\end{align}
\end{subequations}
The equations for the cloud centers and cloud widths (Eqs.\ (\ref{xj_eom_final}), (\ref{yj_eom_final}), (\ref{zj_eom_final}), and (\ref{wjeta_eoms_final})) form a closed set that contain only the $\eta_{j}$, $\dot{\eta}_{j}$, $w_{j\eta}$, and $\dot{w}_{j\eta}$.  Once these quantities are obtained, all of the other variational parameters can be calculated.

The factor $U^{(3D)}({\bf x},{\bf w})$ is the ``variational potential''
\begin{equation}
U^{(3D)}({\bf x},{\bf w}) 
\equiv
U_{\rm ext}^{(3D)}({\bf x},{\bf w}) +
U_{\rm int}^{(3D)}({\bf x},{\bf w}).
\end{equation}
The external and interaction variational potentials are the expectation values of the actual external potential and the condensate density over the trial wave function.  The expression for $U_{\rm ext}^{(3D)}\left({\bf x},{\bf w}\right)$ is 
\begin{subequations}
\begin{eqnarray}
U_{\rm ext}^{(3D)}
\left({\bf x},{\bf w}\right)
&\equiv&
N_{c}
\int_{-\infty}^{\infty}dx\,
\int_{-\infty}^{\infty}dy\,
\int_{-\infty}^{\infty}dz\,
\Psi^{\ast}({\bf r},t)
V_{\rm ext}({\bf r},t)
\Psi({\bf r},t)
\label{genexp}\\
&=&
\sum_{j=1}^{N_{c}}
\left(
\frac{1}{\pi^{3/2}w_{jx}w_{jy}w_{jz}}
\right)
\int\,d^{3}r\,
\exp
\left\{-\sum_{\eta=x,y,z}
\frac
{\left(\eta-\eta_{j}\right)^{2}}
{w_{j\eta}^{2}}
\right\}
V_{\rm ext}({\bf r},t).\label{partexp}
\end{eqnarray}
\end{subequations}
We note here that the variational potential definition given here is smaller than that defined in Ref.\,\cite{PhysRevA.99.043615} by a factor of 2.

The equations of motion for the cloud centers and widths are valid for any external potential. This variational potential is only a function of the center and width parameters of all of the Gaussians.  This model is capable of simulating extreme AI processes such as multiple high-momentum clouds, large volumes, and long interrogation times in a few minutes on a commodity desktop computer.  

\begin{figure}
\begin{center}
\includegraphics[height=2.5in]{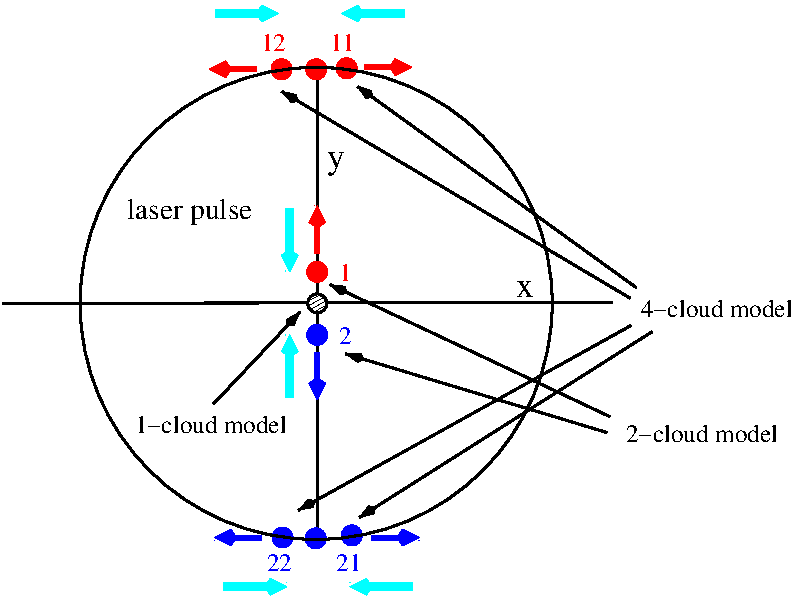}
\caption{The Virginia dual Sagnac interferometer sequence as viewed from a non-rotating frame. A BEC (gray circle) is formed in an ideal harmonic trap ($\omega_{x}=\omega_{y}\equiv\omega_{\perp}$) at the trap center. {\bf First Split:}\,laser pulses are used to split the BEC into two clouds that move at speed $v_{B}$ along the $+y$ axis (cloud 1) and the $-y$ axis (cloud 2), respectively.  {\bf Second Split:}\,At time $t=t_{1}$ cloud 1 at the top is split into clouds 11 and 12. Cloud 11 has a $+v_{B}\hat{\bf i}$ added to its velocity by the split while cloud 12 has $-v_{B}\hat{\bf i}$ added.  These clouds move around a circular orbit in opposite directions. Cloud 2 at the bottom is split into clouds 21 and 22 that also orbit oppositely. {\bf Final split:}\,both of these cloud pairs are allowed to execute one orbit and, at time $t=t_{2}$, when each pair of clouds re-overlaps they are split in the same way as the Second Split. Each re-overlapped pair is split into four clouds: two overlapping clouds that are nearly motionless and two that continue orbiting in opposite directions. Thus the (11,12) cloud pair form one Sagnac interferometer which we will call the ``plus'' (+) Sagnac interferometer and the (21,22) cloud pair forms the ``minus'' (-) Sagnac interferometer.}
\label{sackett_sequence}
\end{center}
\end{figure}

%%%%%%%%%%%%%%%%%%%%%%%%%%%%%%%%%%%%%%%%%%%%%%%%%%
%             Sackett Interferometer             %
%%%%%%%%%%%%%%%%%%%%%%%%%%%%%%%%%%%%%%%%%%%%%%%%%%
\section{The Virginia dual-Sagnac atom-interferometer sequence}
\label{sackett_ai}

To illustrate how this variational model can be used to simulate an actual atom-interferometer sequence we will apply it to a recent dual-Sagnac atom interferometer experiment conducted at the University of Virginia.  The goal of this experiment was to implement a high-precision rotation sensor using a confined BEC~\cite{PhysRevLett.124.120403}.  

The steps of this experiment are described in detail in Figure\,\ref{sackett_sequence} are summarized here as follows. A BEC is formed in a nominally harmonic trap and is initially split by a pair of counter propagating laser pulses into two clouds along the $\pm y$ axis at velocities $\pm v_{B}\hat{\bf j}$ (First Split).  The speed $v_{B}=4\pi\hbar/(M\lambda_{L})$ transfers the momentum of two laser photons to each condensate atom.  

The two clouds fly apart until, at time $t=t_{1}$, they are each split along the $x$ axis to the new clouds, by another pair of laser pulses (Second Split). The time $t_{1}$ is one quarter of the horizontal trap period where the two clouds would be stopped in a purely harmonic trap. The Second Split creates two pairs of counter-orbiting clouds. Each pair of clouds implements a Sagnac interferometer. We will call the pair of clouds shown at the top in the figure (labeled 11 and 12) the plus (+) interferometer and the bottom pair (labeled 21 and 22) the minus (-) interferometer. 

These clouds are allowed to orbit until time $t=t_{2}$ at which time the $x$-axis splitting lasers are applied again (Final Split). If the potential were purely harmonic, the time elapsed since time $t_{1}$ would be one horizontal trap period. As we shall see below the optimal value for the time $t_{2}-t_{1}$ should be the time of maximum overlap of the orbiting clouds which will differ from the trap period due to anharmonic terms present in the external potential. For each Sagnac interferometer this produces a pair of overlapped clouds of stopped atoms plus two clouds that continue orbiting. An image of the system is taken at this point.

Analysis of this image enables the fraction of stopped atoms, the ratio of stopped atoms to total number of atoms in each interferometer, to be measured for this instance of the AI sequence. The result is a stopped-atom fraction for the top interferometer, $S_{+}$, and a fraction for the bottom interferometer, $S_{-}$.  Due to mechanical vibrations and other technical noise sources, the phase measured in each single interferometer is noisy. However, the phase difference between the two interferometers is stable, and nominally given by the Sagnac expression. To extract the Sagnac phase, the AI sequence is repeated a number of times under the same conditions and the values of $S_{+}$ are plotted versus $S_{-}$.  These data points are fitted to an ellipse from which can be extracted the differential phase difference:
\begin{equation}
\Delta\Phi = \Delta\Phi_{+} - \Delta\Phi_{-}.
\end{equation}
where $\Delta\Phi_{\pm}$ is the phase difference between the two stopped-atom clouds in the $\pm$ Sagnac interferometers, respectively.  In this way common-mode noise sources are subtracted out while the Sagnac phase contributions add since they affect the two interferometers oppositely.  Our model accounts for Sagnac effects and for interaction and finite-size effects.  There are no noise sources in the model so we can determine $\Delta\Phi_{\pm}$ for each interferometer separately with no need for ellipse fitting.

Careful analysis of the TOP (time-averaged orbiting potential) trap used in the experiment carried out by the Virginia group indicated that there were anharmonic terms present in the potential that could impact the performance of the interferometer\,\cite{sackett_anharmonic}. The potential experienced by the condensate atoms can be expressed as follows
\begin{equation}
V({\bf r}) = M\omega_{0}^{2}
\left[
\frac{1}{2}\rho^{2} + 
\frac{1}{2}\lambda^{2}z^{2} + 
\frac{1}{3}az^{3} +
\frac{1}{2}b\rho^{2}z +
\frac{1}{4}c\rho^{4} +
\frac{1}{4}fz^{4} + 
\frac{1}{2}h\rho^{2}z^{2}
\right],
\label{anharmonic_potential}
\end{equation}
where $m$ is the mass of a condensate atom and $\omega_{0}$ is given by
\begin{equation*}
\omega_{0} = 
\left(
\frac{7}{64}
\frac{\mu}{M}
\frac{(B_{1}^{\prime})^{2}}{B_{0}}
\right)^{1/2}.
\end{equation*}
and where $\mu$ is the magnetic dipole moment of the trapped state, $B_{0}$ is the TOP trap bias field and $B_{1}^{\prime}$ is the gradient of the quadrupole magnetic field. In the case of ideal bias and gradient fields, the coefficients appearing in the potential can be calculated in terms of $\kappa \equiv B_{1}^{\prime}/B_{0}$ as follows:
\begin{equation}
\lambda^{2} = \frac{8}{7}
\quad
a = \frac{6}{7}\kappa
\quad
b = \frac{9}{14}\kappa
\quad
c = -\frac{237}{3584}\kappa^{2}
\quad
f = \frac{17}{28}\kappa^{2}
\quad
h = \frac{93}{224}\kappa^{2}.
\end{equation}

%%%%%%%%%%%%%%%%%%%%%%%%%%%%%%%%%%%%%%%%%%%%%%%%%%
%              Power-law potential               %
%%%%%%%%%%%%%%%%%%%%%%%%%%%%%%%%%%%%%%%%%%%%%%%%%%
\section{Model equations of motion for a power-law potential}
\label{power_law}

In order to simulate the dual-Sagnac interferometer sequence described above and account for anharmonic terms in the external potential we will derive the variational equations of motion for a 3D power-law potential. We will assume that the potential has the form
\begin{equation}
V_{\rm ext}({\bf r}) \equiv
\sum_{k=1}^{N_{\rm terms}}
C_{p_{x}(k),p_{y}(k),p_{z}(k)}
x^{p_{x}(k)}y^{p_{y}(k)}z^{p_{z}(k)}.
\label{power_law_potential}
\end{equation}
If we include all triples, $(p_{x},p_{y},p_{z})$, of powers such that 
\begin{equation}
p_{x} + p_{y} + p_{z} \le N_{\rm max}
\quad{\rm then}\quad
N_{\rm terms} = 
\frac{1}{6}N_{\rm max}
\left(
N_{\rm max}^{2} + 6N_{\rm max} + 11
\right)
\label{plp}
\end{equation}
excluding the term where all powers are zero.  For $N_{\rm max}=4$, $N_{\rm terms}=34$.

The equations of motion for the general power-law potential are derived in Appendix\,\ref{appA}.  The result is
\begin{subequations}
\begin{align}
\label{xj_eom_final_pl}
\ddot{x}_{j} &=
2\Omega_{z}\dot{y}_{j} +
\Omega_{z}^{2}x_{j} -
\frac{1}{M}
\frac{\partial U_{\rm ext}^{(3D)}}{\partial x_{j}} - 
\frac{1}{M}
F_{jx}({\bf x},{\bf w}),\\
\label{yj_eom_final_pl}
\ddot{y}_{j} &=
-2\Omega_{z}\dot{x}_{j} +
\Omega_{z}^{2}y_{j} -
\frac{1}{M}
\frac{\partial U_{\rm ext}^{(3D)}}{\partial y_{j}} - 
\frac{1}{M}
F_{jy}({\bf x},{\bf w}),\\
\label{zj_eom_final_pl}
\ddot{z}_{j} &=
-\frac{1}{M}
\frac{\partial U_{\rm ext}^{(3D)}}{\partial z_{j}} - 
\frac{1}{M}
F_{jz}({\bf x},{\bf w}),\\
\label{wjeta_eoms_final_pl}
\ddot{w}_{j\eta} &=
\frac{\hbar^{2}}{M^{2}}
w_{j\eta}^{-3} - 
\frac{2}{M}
\frac{\partial U_{\rm ext}^{(3D)}}{\partial w_{j\eta}} +
\frac{gN}{(2\pi)^{3/2}N_{c}M}
\left(
\frac{1}{w_{jx}w_{jy}w_{jz}w_{j\eta}}
\right) -
\frac{1}{M}
W_{j\eta}({\bf x},{\bf w}),\\
\eta &= x,y,z\quad j=1,\dots,N_{c}\nonumber
\end{align}
\end{subequations}
The full expression for $U_{\rm ext}^{(3D)}({\bf x},{\bf w})$ for the general power-law potential is given by Eq.\,(\ref{var_plp}). Expressions for $F_{j\eta}({\bf x},{\bf w})$ and $W_{j\eta}({\bf x},{\bf w})$ were derived in Ref.\,\cite{PhysRevA.99.043615} and are given in full in Appendix\,\ref{appB}.

These equations of motion apply to a system of $N_{c}$ condensate clouds that are subjected to the power-law potential defined above.  The terms $F_{j\eta}({\bf x},{\bf w})$ in the cloud-center equations of motion represent interaction forces exerted on cloud $j$ when it overlaps one or more of the other clouds. The third term on the right-hand-side of Eq.\,(\ref{wjeta_eoms_final_pl}) accounts for the effects of self interaction of cloud $j$. They cause cloud $j$ to expand or contract during its evolution.  The terms $W_{j\eta}({\bf x},{\bf w})$ account for the evolution of the widths of cloud $j$ when it overlaps other clouds.  It is worth noting that the $F_{j\eta}$ and $W_{j\eta}$ terms are negligible when clouds are not overlapping.  

%%%%%%%%%%%%%%%%%%%%%%%%%%%%%%%%%%%%%%%%%%%%%%%%%%
%              Power-law potential               %
%%%%%%%%%%%%%%%%%%%%%%%%%%%%%%%%%%%%%%%%%%%%%%%%%%
\subsection{Virginia trap potential equations of motion}
\label{anharmonic_eoms}

We can derive the equations of motion for the particular case of the anharmonic potential present in the Virginia experiment described earlier and given in Eq.\,(\ref{anharmonic_potential}). This potential can be written in the form of the general power-law potential defined in Eq.\,(\ref{power_law_potential})
\begin{eqnarray}
V_{\rm ext}({\bf r}) 
&=& 
C_{200}x^{2}y^{0}z^{0} +
C_{020}x^{0}y^{2}z^{0} +
C_{002}x^{0}y^{0}z^{2} +
C_{201}x^{2}y^{0}z^{1} +
C_{021}x^{0}y^{2}z^{1} +
C_{003}x^{0}y^{0}z^{3}\nonumber\\
&+&
C_{400}x^{4}y^{0}z^{0} +
C_{220}x^{2}y^{2}z^{0} +
C_{040}x^{0}y^{4}z^{0} +
C_{202}x^{2}y^{0}z^{2} +
C_{022}x^{0}y^{2}z^{2} +
C_{004}x^{0}y^{0}z^{4}.
\label{anharm_pot_scaled}
\end{eqnarray}

Comparing with Eq.\,(\ref{anharmonic_potential}) the coefficients are given by
\begin{equation*}
C_{200} = 
C_{020} = 
C_{002}/\lambda^{2} = \tfrac{1}{2}M\omega_{0}^{2},
\quad
C_{201} = 
C_{021} = \tfrac{1}{2}M\omega_{0}^{2}b,
\quad
C_{003} = \tfrac{1}{3}M\omega_{0}^{2}a
\end{equation*}
\begin{equation*}
C_{400} = 
C_{040} = 
C_{220}/2 = \tfrac{1}{4}M\omega_{0}^{2}c,
\quad
C_{202} = 
C_{022} = \tfrac{1}{2}M\omega_{0}^{2}h,
\quad
C_{004} = \tfrac{1}{4}M\omega_{0}^{2}f
\end{equation*}

We can find the variational potential, $U_{\rm ext}^{(3D)}$, corresponding to the  anharmonic potential given above in Eq.\,(\ref{anharm_pot_scaled}) by using Eq.\,(\ref{uextsc}) for the general power-law potential. Each term in $V_{\rm ext}$ has a corresponding term in $U_{\rm ext}^{(3D)}$.  The term in $U_{\rm ext}^{(3D)}$ corresponding to $C_{\alpha\beta\gamma} x^{\alpha}y^{\beta}z^{\gamma}$ in $V_{\rm ext}$ is $C_{\alpha\beta\gamma}J_{\alpha}(x_{j},w_{jx})J_{\beta}(y_{j},w_{jy})J_{\gamma}(z_{j},w_{jz})$. Since the maximum power is 4 we only need the following $J_{n}(\eta,w)$: $J_{0}=1$, $J_{1}=\eta$, $J_{2}=\eta^{2}+\tfrac{1}{2}w^{2}$, $J_{3}=\eta^{3}+\tfrac{3}{2}\eta w^{2}$, and $J_{4}=\eta^{4}+3\eta^{2}w^{2}+\tfrac{3}{4}w^{4}$.

This process yields the specific form of $U_{\rm ext}^{(3D)}$ for the Virginia potential.  The result is
\begin{eqnarray}
U_{\rm ext}^{(3D)}({\bf x},{\bf w})
&=&
\sum_{j=1}^{N_{\rm terms}}
\Big(
C_{200}
(x_{j}^{2}+\tfrac{1}{2}w_{jx}^{2}) +
C_{020}
(y_{j}^{2}+\tfrac{1}{2}w_{jy}^{2}) +
C_{002}
(z_{j}^{2}+\tfrac{1}{2}w_{jz}^{2})\nonumber\\
&+&
C_{201}
(x_{j}^{2}+\tfrac{1}{2}w_{jx}^{2})z_{j} +
C_{021}
(y_{j}^{2}+\tfrac{1}{2}w_{jy}^{2})z_{j} +
C_{003}
(z_{j}^{3}+\tfrac{3}{2}z_{j}w_{jz}^{2})\nonumber\\
&+&
C_{400}
(x_{j}^{4}+3x_{j}^{2}w_{jx}^{2}+\tfrac{3}{4}w_{jx}^{4}) +
C_{220}
(x_{j}^{2}+\tfrac{1}{2}w_{jx}^{2})
(y_{j}^{2}+\tfrac{1}{2}w_{jy}^{2})\nonumber\\
&+&
C_{040}
(y_{j}^{4}+3y_{j}^{2}w_{jy}^{2}+\tfrac{3}{4}w_{jy}^{4}) +
C_{202}
(x_{j}^{2}+\tfrac{1}{2}w_{jx}^{2})
(z_{j}^{2}+\tfrac{1}{2}w_{jz}^{2})\nonumber\\
&+&
C_{022}
(y_{j}^{2}+\tfrac{1}{2}w_{jy}^{2})
(z_{j}^{2}+\tfrac{1}{2}w_{jz}^{2}) +
C_{004}
(z_{j}^{4}+3z_{j}^{2}w_{jz}^{2}+\tfrac{3}{4}w_{jz}^{4})
\Big).
\label{sackett_variational_potential}
\end{eqnarray}
This potential can be used to write down the equations of motion for the specific case of the Virginia trap potential.

The center-coordinate equations of motion for the potential have the following form:
\begin{eqnarray}
\ddot{x}_{j} + (\omega_{0}^{2}-\Omega_{z}^{2})x_{j}
&=&
+2\Omega_{z}\dot{y}_{j} -
\omega_{0}^{2}bz_{j}x_{j} -
\omega_{0}^{2}c
(x_{j}^{2}+\tfrac{3}{2}w_{jx}^{2})x_{j} -
\omega_{0}^{2}c
(y_{j}^{2}+\tfrac{1}{2}w_{jy}^{2})x_{j}\nonumber\\
&-&
\omega_{0}^{2}h
(z_{j}^{2}+\tfrac{1}{2}w_{jz}^{2})x_{j} -
\frac{1}{M}
F_{jx}({\bf x},{\bf w})\nonumber\\
\ddot{y}_{j} + (\omega_{0}^{2}-\Omega_{z}^{2})y_{j}
&=&
-2\Omega_{z}\dot{x}_{j} -
\omega_{0}^{2}bz_{j}y_{j} -
\omega_{0}^{2}c
(y_{j}^{2}+\tfrac{3}{2}w_{jy}^{2})y_{j} -
\omega_{0}^{2}c
(x_{j}^{2}+\tfrac{1}{2}w_{jx}^{2})y_{j}\nonumber\\
&-&
\omega_{0}^{2}h
(z_{j}^{2}+\tfrac{1}{2}w_{jz}^{2})y_{j} -
\frac{1}{M}
F_{jy}({\bf x},{\bf w})\nonumber\\
\ddot{z}_{j} + \lambda^{2}\omega_{0}^{2}z_{j}
&=&
-\tfrac{1}{2}\omega_{0}^{2}b
(x_{j}^{2}+\tfrac{1}{2}w_{jx}^{2}) -
\tfrac{1}{2}\omega_{0}^{2}b
(y_{j}^{2}+\tfrac{1}{2}w_{jy}^{2}) -
\omega_{0}^{2}a
(z_{j}^{2}+\tfrac{1}{2}w_{jz}^{2})\nonumber\\
&-&
\omega_{0}^{2}h
(x_{j}^{2}+\tfrac{1}{2}w_{jx}^{2})z_{j} -
\omega_{0}^{2}h
(y_{j}^{2}+\tfrac{1}{2}w_{jy}^{2})z_{j} -
\omega_{0}^{2}f
(z_{j}^{2}+\tfrac{3}{2}w_{jz}^{2})z_{j} -
\frac{1}{M}
F_{jz}({\bf x},{\bf w})\nonumber\\
j &=& 1,\dots,N_{c}
\label{full_center_eoms}
\end{eqnarray}
The EOMs for the widths are
\begin{eqnarray}
\ddot{w}_{jx} +
\omega_{0}^{2}w_{jx}
&=&
\frac{\hbar^{2}}{M^{2}}w_{jx}^{-3} +
\frac{gN}{(2\pi)^{3/2}N_{c}M}
\left(
\frac{1}{w_{jx}w_{jy}w_{jz}w_{jx}}
\right) -
\omega_{0}^{2}bz_{j}w_{jx} -
3\omega_{0}^{2}c
(x_{j}^{2}+\tfrac{1}{2}w_{jx}^{2})w_{jx}\nonumber\\
&-&
\omega_{0}^{2}c
(y_{j}^{2}+\tfrac{1}{2}w_{jy}^{2})w_{jx} -
\omega_{0}^{2}h
(z_{j}^{2}+\tfrac{1}{2}w_{jz}^{2})w_{jx} - 
\frac{1}{M}
W_{jx}({\bf x},{\bf w})\nonumber\\
\ddot{w}_{jy} +
\omega_{0}^{2}w_{jy}
&=&
\frac{\hbar^{2}}{M^{2}}w_{jy}^{-3} +
\frac{gN}{(2\pi)^{3/2}N_{c}M}
\left(
\frac{1}{w_{jx}w_{jy}w_{jz}w_{jy}}
\right) -
\omega_{0}^{2}bz_{j}w_{jy} -
3\omega_{0}^{2}c
(y_{j}^{2}+\tfrac{1}{2}w_{jy}^{2})w_{jy}\nonumber\\
&-&
\omega_{0}^{2}c
(x_{j}^{2}+\tfrac{1}{2}w_{jx}^{2})w_{jy} -
\omega_{0}^{2}h
(z_{j}^{2}+\tfrac{1}{2}w_{jz}^{2})w_{jy} - 
\frac{1}{M}
W_{jy}({\bf x},{\bf w})\nonumber\\
\ddot{w}_{jz} +
\lambda^{2}\omega_{0}^{2}w_{jz}
&=&
\frac{\hbar^{2}}{M^{2}}w_{jz}^{-3} +
\frac{gN}{(2\pi)^{3/2}N_{c}M}
\left(
\frac{1}{w_{jx}w_{jy}w_{jz}w_{jz}}
\right) -
2\omega_{0}^{2}az_{j}w_{jz} -
3\omega_{0}^{2}f
(z_{j}^{2}+\tfrac{1}{2}w_{jz}^{2})w_{jz}\nonumber\\
&-&
\omega_{0}^{2}h
(x_{j}^{2}+\tfrac{1}{2}w_{jx}^{2})w_{jz} -
\omega_{0}^{2}h
(y_{j}^{2}+\tfrac{1}{2}w_{jy}^{2})w_{jz} - 
\frac{1}{M}
W_{jz}({\bf x},{\bf w})\nonumber\\
j &=& 1,\dots,N_{c}
\label{full_width_eoms}
\end{eqnarray}
We again note here that the terms $F_{j\eta}({\bf x},{\bf w})$ and $W_{j\eta}({\bf x},{\bf w})$ account for interactions between different clouds and couple the widths and center coordinates of cloud $j$ to the widths and centers of all the other clouds. These terms are negligible unless cloud $j$ has a spatial overlap with another cloud.

%%%%%%%%%%%%%%%%%%%%%%%%%%%%%%%%%%%%%%%%%%%%%%%%%%%%%%%%%%%%%
% Applying the variational model to the Sackett AI sequence %
%%%%%%%%%%%%%%%%%%%%%%%%%%%%%%%%%%%%%%%%%%%%%%%%%%%%%%%%%%%%%

\section{Computing the stopped-atom fraction within the variational model}
\label{stopped_atom_fraction}

Our model is applied to the Virginia interferometer sequence by first finding stationary values of the center coordinates and $x$,$y$,$z$ widths in the one-cloud ($N_{c}=1$) version of the equations given in Eqs.\,(\ref{xj_eom_final})-(\ref{betajeta_eoms_final}) to model the initial condensate. The two-cloud ($N_{c}=2$) version of the model is used to simulate the trajectories after the First Split.  The one-cloud-model values for the center coordinates and widths plus their $y$ velocities modified by the kick imparted by the lasers in the First Split are used as initial conditions for the two-cloud model.  Following the protocol, after one-quarter of the trap period, the Second Split is applied and the four-cloud version ($N_{c}=4$) of the equations is used to simulate the system evolution for one full trap period. The final values of the centers and widths and their derivatives (modified to include the laser kick applied in the Second Split) are used as the initial conditions for the four-cloud model.  

The variational trial wave function given by Eq.\,(\ref{trial_wf_3d}), at time $t=t_{2}$ (four-cloud model) just before the Final Split, can be written as
\begin{eqnarray*}
\Psi({\bf r},t_{2}) 
&=& 
\frac{1}{2}
\left(
A_{11}(t)e^{g_{11}({\bf r},t_{2})} +
A_{12}(t)e^{g_{12}({\bf r},t_{2})} +
A_{21}(t)e^{g_{21}({\bf r},t_{2})} +
A_{22}(t)e^{g_{22}({\bf r},t_{2})}
\right)\\
&\equiv&
\psi_{11}({\bf r},t_{2}) +
\psi_{12}({\bf r},t_{2}) +
\psi_{21}({\bf r},t_{2}) +
\psi_{22}({\bf r},t_{2}).
\end{eqnarray*}
where
\begin{eqnarray*}
g_{ij}({\bf r},t_{2})
&\equiv&
\sum_{\eta=x,y,z}
\Bigg(
-\frac{(\eta-\eta_{ij}(t_{2}))^{2}}
{2(w_{ij\eta}(t_{2}))^{2}} +
i\big(\epsilon_{ij\eta}(t_{2})\eta+
\beta_{ij}(t_{2})\eta^{2}\big)
\Bigg),
\quad
ij = 11,12,21,22
\end{eqnarray*}
and
\begin{eqnarray*}
\epsilon_{ijx}(t_{2})
&\equiv&
\frac{M}{\hbar}
\left(
\dot{x}_{ij}(t_{2}) -
\Omega_{z}y_{ij}(t_{2})
\right) - 
2\beta_{ijx}(t_{2})
x_{ij}(t_{2})\\
\epsilon_{ijy}(t_{2})
&\equiv&
\frac{M}{\hbar}
\left(
\dot{y}_{ij}(t_{2}) + 
\Omega_{z}x_{ij}(t_{2}) 
\right) - 
2\beta_{ijy}(t_{2})
y_{ij}(t_{2})\\
\epsilon_{ijz}(t_{2})
&\equiv&
\frac{M}{\hbar}
\dot{z}_{ij}(t_{2}) - 
2\beta_{ijz}(t_{2})
z_{ij}(t_{2}).
\end{eqnarray*}
The first two terms, $\psi_{11}$ and $\psi_{12}$, represent the plus (top) Sagnac interferometer shown in Fig.\,\ref{sackett_sequence} while the last two terms, $\psi_{21}$ and $\psi_{22}$, represent the minus (bottom) Sagnac interferometer.  We note that clouds belonging to different interferometers at time $t=t_{2}$ have no spatial overlap so that any product of a plus interferometer cloud and a minus cloud will be strictly zero.

The final split has the effect of stopping half of cloud {\it ij} and leaving the other half alone ({\it ij}=11,12,21, or 22).  Since, at $t=t_{2}$, the clouds 11 and 21 move nearly along $+x$ and clouds 12 and 22 move nearly along $-x$, after the final split we have
\begin{equation*}
\psi_{ij}({\bf r},t_{2})
\rightarrow
\frac{1}{\sqrt{2}}
\left(
\psi_{ij}({\bf r},t_{2}) + 
\psi_{ij}({\bf r},t_{2})
e^{-i\lambda_{ij}Mv_{B}x/\hbar}
\right)
\end{equation*}
where $\lambda_{11}=\lambda_{21}=-\lambda_{12}=-\lambda_{22}=1$ and
where the term with the exponential is the stopped cloud. 
Since there is no spatial overlap between any linear combination of the 11 and 12 clouds with any linear combination of the 21 and 22 clouds, we may treat them separately.  Each constitutes a separate Sagnac interferometer.

The quantities that can be determined from the image data generated in the Virginia experiment are the ratios, $S_{\pm}$, of the number of stopped atoms to the total number of atoms in the two Sagnac interferometers.  We can compute this from the variational wave function just at the moment of the final split as the probability of being in the zero-momentum state.  Since the clouds of the two interferometers have no spatial overlap, the fraction of stopped atoms in each interferometer can be written as
\begin{equation*}
S_{+} = 
\int\,d^{3}r
\left|
\psi_{11}({\bf r},t_{2})
e^{-iMv_{B}x/\hbar} +
\psi_{12}({\bf r},t_{2})
e^{+iMv_{B}x/\hbar}
\right|^{2}
\end{equation*}
and
\begin{equation*}
S_{-} = 
\int\,d^{3}r
\left|
\psi_{21}({\bf r},t_{2})
e^{-iMv_{B}x/\hbar} +
\psi_{22}({\bf r},t_{2})
e^{+iMv_{B}x/\hbar}
\right|^{2}
\end{equation*}
We can express $S_{\pm}$ in terms of the variational parameters by inserting the expression for the trial wave function into the above integrals and carrying out the integration.  The result is
\begin{eqnarray}
S_{+}
&=&
\frac{1}{2} +
\frac{1}{2}
{\rm Re}
\Big\{
M(
x_{11}(t_{2}),
w_{11x}(t_{2}),
\epsilon_{11x}(t_{2}),
\beta_{11x}(t_{2}),
x_{12}(t_{2}),
w_{12x}(t_{2}),
\epsilon_{12x}(t_{2})+\tfrac{Mv_{B}}{\hbar},
\beta_{12x}(t_{2}))\nonumber\\
&\times&
M(
y_{11}(t_{2}),
w_{11y}(t_{2}),
\epsilon_{11y}(t_{2}),
\beta_{11y}(t_{2}),
y_{12}(t_{2}),
w_{12y}(t_{2}),
\epsilon_{12y}(t_{2}),
\beta_{12y}(t_{2}))\nonumber\\
&\times&
M(
z_{11}(t_{2}),
w_{11z}(t_{2}),
\epsilon_{11z}(t_{2}),
\beta_{11z}(t_{2}),
z_{12}(t_{2}),
w_{12z}(t_{2}),
\epsilon_{12z}(t_{2}),
\beta_{12z}(t_{2}))
\Big\}\label{Sp_exact}\\
S_{-}
&=&
\frac{1}{2} +
\frac{1}{2}
{\rm Re}
\Big\{
M(
x_{21}(t_{2}),
w_{21x}(t_{2}),
\epsilon_{21x}(t_{2}),
\beta_{21x}(t_{2}),
x_{22}(t_{2}),
w_{22x}(t_{2}),
\epsilon_{22x}(t_{2})+\tfrac{Mv_{B}}{\hbar},
\beta_{22x}(t_{2}))\nonumber\\
&\times&
M(
y_{21}(t_{2}),
w_{21y}(t_{2}),
\epsilon_{21y}(t_{2}),
\beta_{21y}(t_{2}),
y_{22}(t_{2}),
w_{22y}(t_{2}),
\epsilon_{22y}(t_{2}),
\beta_{22y}(t_{2}))\nonumber\\
&\times&
M(
z_{21}(t_{2}),
w_{21z}(t_{2}),
\epsilon_{21z}(t_{2}),
\beta_{21z}(t_{2}),
z_{22}(t_{2}),
w_{22z}(t_{2}),
\epsilon_{22z}(t_{2}),
\beta_{22z}(t_{2}))
\Big\}
\label{Sm_exact}
\end{eqnarray}
where
\begin{eqnarray}
M(
x_{1},w_{1},\epsilon_{1},\beta_{1},
x_{2},w_{2},\epsilon_{2},\beta_{2})
&=&
\frac
{\exp
\Bigg\{
\frac
{\big(
\frac{x_{1}}{2w_{1}^{2}} +
\frac{x_{2}}{2w_{2}^{2}} +
\tfrac{1}{2}i
(\epsilon_{2}-\epsilon_{1})
\big)^{2}}
{\big(
\frac{1}{2w_{1}^{2}} +
\frac{1}{2w_{2}^{2}} -
i(\beta_{2}-\beta_{1})
\big)} -
\Big(
\frac{x_{1}^{2}}{2w_{1}^{2}} +
\frac{x_{2}^{2}}{2w_{2}^{2}}
\Big)
\Bigg\}}
{\Big(
\frac{w_{2}}{2w_{1}} +
\frac{w_{1}}{2w_{2}} -
i(\beta_{2}-\beta_{1})w_{1}w_{2}
\Big)^{1/2}}.
\label{m_factor}
\end{eqnarray}
These expressions enable us to compute $S_{\pm}$ in terms of the values of the variational parameters just before the Final Split.

%%%%%%%%%%%%%%%%%%%%%%%%%%%%%%%%%%%%%%%%%%%%%%%%%%%%%%%%%%%%%
%         zero-rotation stopped-atom fraction           %
%%%%%%%%%%%%%%%%%%%%%%%%%%%%%%%%%%%%%%%%%%%%%%%%%%%%%%%%%%%%%
\section{Approximate expressions for the stopped-atom fraction}
\label{approx_solutions}
\subsection{Stopped-atom fraction for zero rotation speed}
\label{zero_rotation_S_plus}

We can derive an approximate expression for $S_{+}$ for the case of the Virginia trap potential, where cloud-cloud interactions are neglected, and where the rotating-frame speed is zero ($\Omega_{z}=0$).  We can simplify the formula for $S_{+}(\Omega_{z})$ (Eq.\,(\ref{Sp_exact})) at $\Omega_{z}=0$ by taking advantage of certain symmetries in the solution of the model equations for the zero-rotation case.  These are that the $x$ and $\dot{x}$ of clouds 11 and 12 are negatives of each other, the $y$ and $\dot{y}$ of these clouds are the same, and the $z$ and $\dot{z}$ of the clouds are the same at all times.  Quantitatively we have
\begin{equation*}
x_{11}(t)\equiv x_{1}(t),
\quad
x_{12}(t)\equiv x_{2}(t),
\quad
x_{1}(t) = - x_{2}(t),
\quad
\dot{x}_{1}(t) = - \dot{x}_{2}(t)
\end{equation*}
\begin{equation*}
y_{11}(t) = y_{12}(t),
\quad
\dot{y}_{11}(t) = \dot{y}_{12}(t),
\quad
z_{11}(t) = z_{12}(t),
\quad
\dot{z}_{11}(t) = \dot{z}_{12}(t),
\end{equation*}

Furthermore all of the $x$ and $y$ widths and their dots for both clouds are the same at all times:
\begin{equation*}
w_{11x}(t) = 
w_{12x}(t) = 
w_{11y}(t) =
w_{12y}(t) \equiv 
w_{\perp}(t)
\end{equation*}
\begin{equation*}
\dot{w}_{11x}(t) = 
\dot{w}_{12x}(t) = 
\dot{w}_{11y}(t) =
\dot{w}_{12y}(t) \equiv 
\dot{w}_{\perp}(t)
\end{equation*}
Also, the $z$ widths and dot widths of the two clouds are equal
\begin{equation*}
w_{11z}(t) = 
w_{12z}(t)  \equiv 
w_{z}(t),
\quad
{\rm and}
\quad
\dot{w}_{11z}(t) = 
\dot{w}_{12z}(t)  \equiv 
\dot{w}_{z}(t)
\end{equation*}
These symmetries greatly simplify the expression for $S_{+}(0)$

The simplified expression is
\begin{equation}
S_{+}(0) =
\tfrac{1}{2}
+ 
\tfrac{1}{2}
\exp
\Bigg\{
-\left(\frac{M}{\hbar}\right)^{2}
\left[
\textcolor{black}{
\tfrac{1}{2}
\big(
2v_{B} -
(\dot{x}_{1} - \dot{x}_{2})
\big)w_{\perp}} +
\textcolor{black}{
(x_{1} - x_{2})
\dot{w}_{\perp}}
\right]^{2} -
\Big(
\textcolor{black}{
\frac{x_{1}^{2}}{2w_{\perp}^{2}} +
\frac{x_{2}^{2}}{2w_{\perp}^{2}}
}
\Big)
\Bigg\}
\label{sp_formula_SI}
\end{equation}
where $x_{1}$, $x_{2}$, $\dot{x}_{1}$, $\dot{x}_{2}$, $w_{\perp}$, and $\dot{w}_{\perp}$ are all evaluated at time $t=t_{2}$, the moment of the final split. This result provides insight into the model-predicted physical mechanisms that affect the fraction of stopped atoms when the counter-orbiting clouds recombine just before the final split.

The exponential in Eq.\,(\ref{sp_formula_SI}) contains two terms.  The first term, shown enclosed by square brackets, is due to the relative velocity of the two clouds.  The second term, shown in parenthesis, is due to the finite widths of the two clouds compared with the separation of their center coordinates.  

The relative velocity term has two contributions.  The first term inside the square brackets is the difference between the relative velocity of the center coordinates of the two clouds, ($\dot{x}_{1}-\dot{x}_{2}$) and their initial relative velocity at the second split ($2v_{B}$).  The second contribution to the relative velocity is due to the expansion or contraction of the two clouds.  Even though the expansion and contraction of the two clouds is the same, if their centers are offset, then there will be a relative velocity due to the time rate of change of the width, $\dot{w}_{\perp}$.

The finite-width term decreases the fraction of stopped atoms the further apart the two clouds are and the narrower their widths are relative to their separation.  Thus the closer together and wider the clouds are, the larger the fraction of stopped atoms.  This is an entirely intuitive result since both decreased separation and larger widths tend to maximize overlap and thus increase the possibility of interference.

%%%%%%%%%%%%%%%%%%%%%%%%%%%%%%%%%%%%%%%%%%%%%%%%%%%%%%%%%%%%%
%                    Harmonic Solution                      %
%%%%%%%%%%%%%%%%%%%%%%%%%%%%%%%%%%%%%%%%%%%%%%%%%%%%%%%%%%%%%
\subsection{Exact stopped-atom fraction for non-interacting clouds in a harmonic potential}
\label{harmonic_solution}

The variational equations of motion applied to the Virginia experiment, found in Eqs.\,(\ref{full_center_eoms}) and (\ref{full_width_eoms}), for the case of a harmonic potential and where cloud-cloud interactions are neglected have the form
\begin{eqnarray}
\ddot{x}_{j} + (\omega_{0}^{2}-\Omega_{z}^{2})x_{j}
&=&
+2\Omega_{z}\dot{y}_{j}\nonumber\\
\ddot{y}_{j} + (\omega_{0}^{2}-\Omega_{z}^{2})y_{j}
&=&
-2\Omega_{z}\dot{x}_{j}\nonumber\\
\ddot{z}_{j} + \lambda^{2}\omega_{0}^{2}z_{j}
&=&
0\nonumber\\
\ddot{w}_{jx} +
\omega_{0}^{2}w_{jx}
&=&
\frac{\hbar^{2}}{M^{2}}w_{jx}^{-3} +
\frac{gN}{(2\pi)^{3/2}N_{c}M}
\left(
\frac{1}{w_{jx}w_{jy}w_{jz}w_{jx}}
\right)\nonumber\\
\ddot{w}_{jy} +
\omega_{0}^{2}w_{jy}
&=&
\frac{\hbar^{2}}{M^{2}}w_{jy}^{-3} +
\frac{gN}{(2\pi)^{3/2}N_{c}M}
\left(
\frac{1}{w_{jx}w_{jy}w_{jz}w_{jy}}
\right)\nonumber\\
\ddot{w}_{jz} +
\lambda^{2}\omega_{0}^{2}w_{jz}
&=&
\frac{\hbar^{2}}{M^{2}}w_{jz}^{-3} +
\frac{gN}{(2\pi)^{3/2}N_{c}M}
\left(
\frac{1}{w_{jx}w_{jy}w_{jz}w_{jz}}
\right)\nonumber\\
j &=& 1,\dots,N_{c}.
\label{harmonic_eoms}
\end{eqnarray}
The equations for the center coordinates of each cloud form a closed set and can be solved exactly.  The width equations must be solved numerically but exhibit a clear symmetry in that the equations for $w_{jx}(t)$ and $w_{jy}(t)$ are the same.  When the initial conditions for these widths are the same (as is the case when applying the one, two, and four-cloud models to the Virginia AI sequence) the solutions $w_{jx}(t)$ and $w_{jy}(t)$ will be identical.  
Note that the evolution of the widths do not depend on the speed, $\Omega_{z}$, of the rotating frame as this does not appear in the width equations of motion.

Using the analytical solutions of the cloud-center equations of motion and the symmetries of the solutions of the width equations of motion we can follows the steps of the Virginia AI sequence to obtain an analytical expression for $S_{+}(\Omega_{z})$. The result is
\begin{equation}
S_{+}(\Omega_{z}) = 
\frac{1}{2} + 
\frac{1}{2}
\exp
\left\{
-\left(
\frac{2Mv_{B}w_{\perp}}{\hbar}
\right)^{2}
\sin^{2}\left(\pi\frac{\Omega_{z}}{\omega_{0}}\right)
\right\}
\cos
\left\{
\left(
\frac{2Mv_{B}^{2}}{\hbar\omega_{0}}
\right)
\left(
\sin\left(\tfrac{5\pi}{2}\tfrac{\Omega_{z}}{\omega_{0}}\right) -
\sin\left(\tfrac{\pi}{2}\tfrac{\Omega_{z}}{\omega_{0}}\right)
\right)
\right\}
\label{Splus}
\end{equation}
and the expression for $S_{-}(\Omega_{z})$ is identical.  In the above $w_{\perp} = w_{jx}(t_{2})=w_{jy}(t_{2})$ is the transverse width of the condensate at the moment of the Final Split.

In the limit where the rotation speed is much smaller than the trap frequency, $\Omega_{z}\ll\omega_{0}$, this approximates to
\begin{equation}
S_{+}(\Omega_{z}) = 
\frac{1}{2} + 
\frac{1}{2}
\exp
\left\{
-\left(
\frac{2\pi M v_{B} w_{\perp}}{\hbar\omega_{0}}
\right)^{2}
\Omega_{z}^{2}
\right\}
\cos
\left\{
\left(
\frac{4M\pi (v_{B}/\omega_{0})^{2}\Omega_{z}}{\hbar}
\right)
\right\}
\label{Splus_approx}
\end{equation}
For a purely harmonic potential the radius of the circular orbit followed by the condensate clouds is given by $R=v_{B}/\omega_{0}$ so that the area of this orbit (and thus the interferometer area) is $A=\pi R^{2}=\pi(v_{B}/\omega_{0})^{2}$.  This enables us to recognize the argument of the cosine above as the Sagnac phase for a single interferometer:
\begin{equation*}
\Phi_{S} = 
\frac{4M\pi (v_{B}/\omega_{0})^{2}\Omega_{z}}{\hbar} = 
\frac{4M A\Omega_{z}}{\hbar}.
\end{equation*}
The expression for $S_{+}(\Omega_{z})$ in Eq.\,(\ref{Splus_approx}) provides guidance for simulations where inter-cloud interactions and anharmonic terms in the potential are included.

\clearpage
\begin{figure}[htb]
\centering
\includegraphics[scale=0.3]{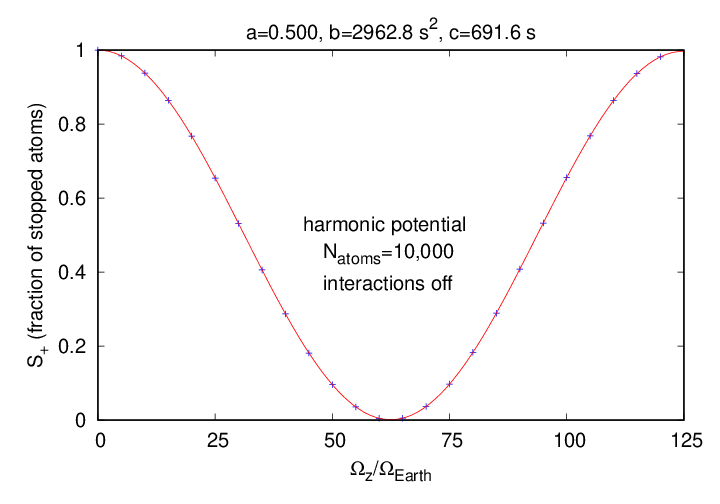}
\includegraphics[scale=0.3]{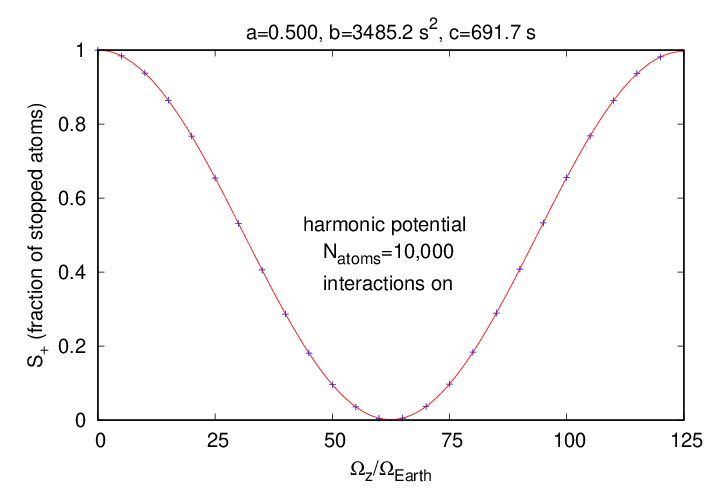}\\
\includegraphics[scale=0.3]{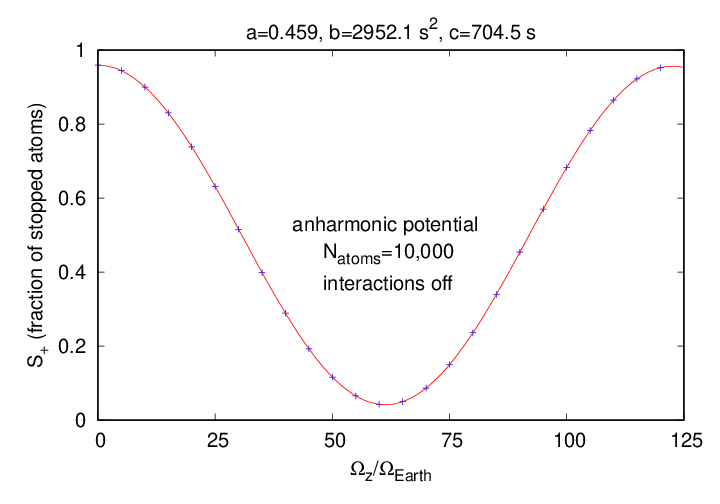}
\includegraphics[scale=0.3]{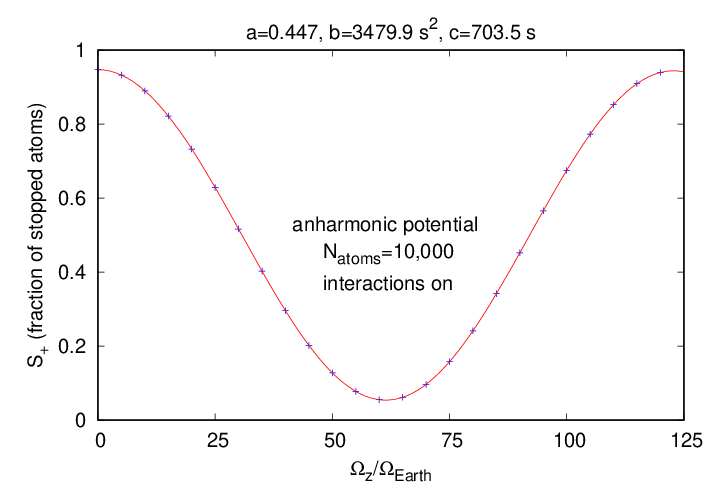}
\caption{The fraction of stopped atoms, $S_{+}$, vs rotating frame speed, $\Omega_{z}$, with $N_{\rm atoms} =$\,10,000 $^{87}$Rb atoms for four different cases. Upper Left: harmonic potential and cloud-cloud interactions off; Upper Right: harmonic potential and interactions on; Lower Left: anharmonic potential and interactions off; Lower Right: anharmonic potential and interactions on.}
\label{ha_001}
\end{figure}

%%%%%%%%%%%%%%%%%%%%%%%%%%%%%%%%%%%%%%%%%%%%%%%%%%%%%%%%%%%%%
%          Interaction and Anharmonic Effect Study          %
%%%%%%%%%%%%%%%%%%%%%%%%%%%%%%%%%%%%%%%%%%%%%%%%%%%%%%%%%%%%%
\section{Interaction and Anharmonic Effect Study}
\label{sim_study}

We studied the effects of interactions and the presence of anharmonic terms in the potential by computing the dependence of $S_{+}$ on the true rotation speed $\Omega_{z}$ by
simulating the interferometer experiment for various conditions.  These conditions included harmonic or anharmonic potential, cloud-cloud interactions on or off, and number of condensate atoms. These numbers were $N=1\times 10^{4}, 1\times 10^{5}, 2\times 10^{5}, 3\times 10^{5}, 4\times 10^{5}, 5\times 10^{5}, 6\times 10^{5}, 7\times 10^{5}, 8\times 10^{5}, 9\times 10^{5}, 1\times 10^{6}, 2\times 10^{6}$ atoms.  In all there were 48 sets of combinations of harmonic or anharmonic potential, cloud-cloud interactions off or on, and number of condensate atoms.  For each of these 48 combinations of potential, interactions, and condensate number we simulated the interferometer experiment at 26 different input rotation speeds at equal intervals ranging from zero up to 125 times the Earth's rotation speed.  The values of $S_{+}$ were then plotted versus $\Omega_{z}$ for each case.  All other parameter values were taken from the original Virginia experiment.

We fitted the $S_{+}$ vs $\Omega_{z}$ plots to a function which has the same dependence on $\Omega_{z}$ as the expression for $S_{+}$ in Eq.\,(\ref{Splus_approx}) for the case of a harmonic potential and no cloud-cloud interactions.  The fit function was
\begin{equation}
S_{+}(a,b,c,\Omega_{z}) = 
\frac{1}{2} + a e^{-b\Omega_{z}^{2}}\cos(2\pi c\Omega_{z}).
\label{fit_function}
\end{equation}
The fits were performed for data where the rotation speeds were expressed in Hertz.  Thus the $b$ parameter is measured in seconds squared and the $c$ parameter is measured in seconds.

\begin{figure}[tb]
\centering
\includegraphics[scale=0.3]{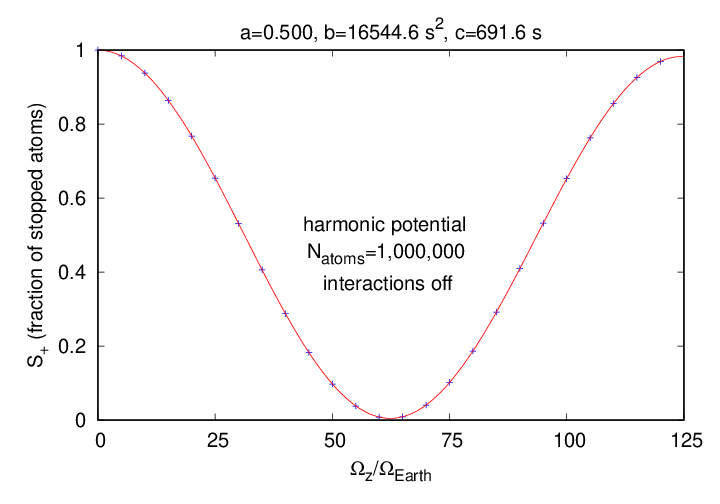}
\includegraphics[scale=0.3]{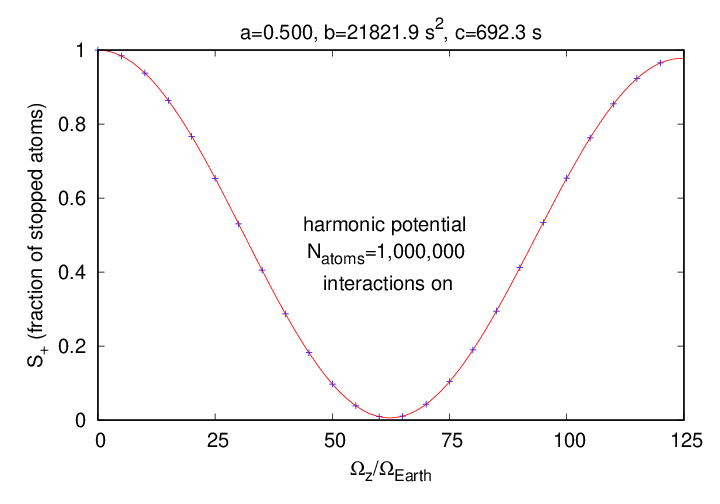}\\
\includegraphics[scale=0.3]{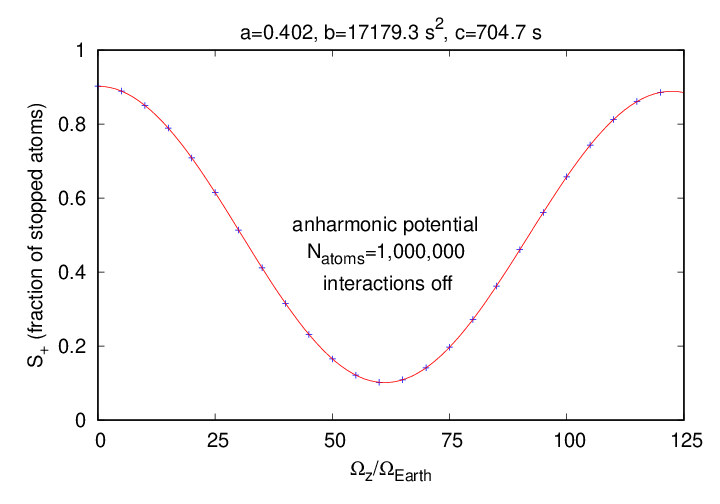}
\includegraphics[scale=0.3]{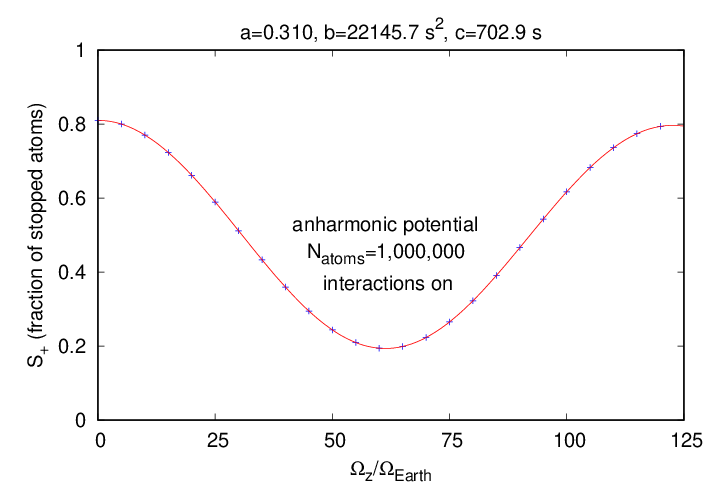}
\caption{The fraction of stopped atoms, $S_{+}$, vs rotating frame speed, $\Omega_{z}$, with $N_{\rm atoms} =$\,1,000,000 $^{87}$Rb atoms for four different cases. Upper Left: harmonic potential and cloud-cloud interactions off; Upper Right: harmonic potential and interactions on; Lower Left: anharmonic potential and interactions off; Lower Right: anharmonic potential and interactions on.}
\label{ha_100}
\end{figure}

We can use Eq.\,(\ref{sp_formula_SI}) to find an approximate expression for the $a$ parameter. Note that, when $\Omega_{z}=0$, the above expression becomes $S_{+}(0)=1/2 + a$.  Comparing this with Eq.\,(\ref{sp_formula_SI}) gives an approximate formula for $a$
\begin{equation}
a \approx
\tfrac{1}{2}
\exp
\Bigg\{
-\left(\frac{M}{\hbar}\right)^{2}
\left[
\textcolor{black}{
\tfrac{1}{2}
\big(
2v_{B} -
(\dot{x}_{1} - \dot{x}_{2})
\big)w_{\perp}} +
\textcolor{black}{
(x_{1} - x_{2})
\dot{w}_{\perp}}
\right]^{2} -
\Big(
\textcolor{black}{
\frac{x_{1}^{2}}{2w_{\perp}^{2}} +
\frac{x_{2}^{2}}{2w_{\perp}^{2}}
}
\Big)
\Bigg\}.
\label{a_formula}
\end{equation}
Comparing the fit function with Eq.\,(\ref{Splus_approx}) gives an approximate expression for the $b$ parameter
\begin{equation}
b \approx 
\left(
\frac{4\pi^{2} M v_{B}}{\hbar\omega_{0}}
\right)^{2} w_{\perp}^{2}
\label{b_formula}
\end{equation}
where we note that this equation gives the value of $b$ when the $\Omega_{z}$ appearing in the exponential is measured in Hz.  

It is clear from these expressions that the width of the clouds, $w_{\perp}$, and to a lesser extent the width velocity, $\dot{w}_{\perp}$, at the final split have a major effect on the value of $S_{+}$.

\clearpage

\begin{table}[htb]
\centering
\begin{tabular}{||c c c c c c c||}
\hline
$N_{atoms}$ & $a$ & $a$ & $b$\ (s$^{2}$) & $b$\ (s$^{2}$) & $c$\ (s) & $c$\ (s)\\
\hline
&(formula) & (fit) & (formula) & (fit) & (formula) & (fit)\\
\hline\hline
 10000  & 0.500 & 0.500 &  2962.6 &  2962.8 & 691.6 & 691.6\\
 100000 & 0.500 & 0.500 &  6733.3 &  6733.6 & 691.6 & 691.6\\
 200000 & 0.500 & 0.500 &  8787.9 &  8788.3 & 691.6 & 691.6\\
 500000 & 0.500 & 0.500 & 12577.9 & 12578.5 & 691.6 & 691.6\\
1000000 & 0.500 & 0.500 & 16543.9 & 16544.6 & 691.6 & 691.6\\
2000000 & 0.500 & 0.500 & 21789.6 & 21790.4 & 691.6 & 691.6\\
\hline
 10000  & 0.459 & 0.459 &  3027.2 &  2952.1 & 702.9 & 704.5\\
 100000 & 0.450 & 0.450 &  6984.8 &  6913.5 & 702.9 & 704.5\\
 200000 & 0.441 & 0.441 &  9137.4 &  9072.4 & 702.9 & 704.6\\
 500000 & 0.422 & 0.422 & 13101.1 & 13039.0 & 702.9 & 704.6\\
1000000 & 0.402 & 0.402 & 17244.1 & 17179.3 & 702.9 & 704.7\\
2000000 & 0.376 & 0.376 & 22720.7 & 22648.1 & 703.0 & 704.7\\
\hline\hline
\end{tabular}
\caption{Comparison of the fitted values of the parameters, $a$, $b$, and $c$ with their values predicted by Eqs.\,(\ref{a_formula}), (\ref{b_formula}), and (\ref{c_formula}) when cloud-cloud interactions are turned off.  The top half of the table gives results for a harmonic potential and the bottom half for anharmonic potential.}
\label{table_int_off}
\end{table}

The $c$ parameter can be approximated by assuming that the quantity appearing in the cosine term is just the Sagnac phase. Since the full potential including the anharmonic terms has cylindrical symmetry, the $z$ component of the angular momentum of each cloud is conserved.  Thus, if $\Omega_{z}$ in the $\cos(2\pi c\Omega_{z})$ factor in Eq.\,(\ref{fit_function}) is measured in Hz, we have
\begin{eqnarray}
c 
&=&
\frac{1}{\hbar}
\int_{t_{i}}^{t_{f}} dt\,
(L_{2z}(t) - L_{1z}(t)) = 
(L_{2z}(t_{f}) - L_{1z}(t_{f}))\int_{t_{i}}^{t_{f}} dt\nonumber\\
c
&\approx&
\left(
\frac{M(t_{f}-t_{i})}{\hbar}
\right)
\left(
(x_{2}\dot{y}_{2}-y_{2}\dot{x}_{2}) -
(x_{1}\dot{y}_{1}-y_{1}\dot{x}_{1})
\right).
\label{c_formula}
\end{eqnarray}
In the last expression all quantities are evaluated at time $t=t_{2}$. It will be instructive to compare the values of $a$, $b$, and $c$ given by these approximate formulas with their values determined from the simulation fits.

Plots of $S_{+}$ versus $\Omega_{z}$ for $N_{\rm atoms} = $\, 10,000 atoms are shown in Figure\,\ref{ha_001} and for $N_{\rm atoms} = $\, 1,000,000 atoms in Figure\,\ref{ha_100}.  The four plots found in these figures are for a harmonic potential with cloud-cloud interactions turned off (upper left), harmonic with interactions on (upper right), anharmonic with interactions off (lower left), and anharmonic with interactions on (lower right). These plots also display the values of $S_{+}(\Omega_{z})$ fitted to the function shown in Eq.\,(\ref{fit_function}).  We note that these fits closely follow the simulation data in all cases.  

\clearpage

\begin{table}[tb]
\centering
\begin{tabular}{||c c c c c c c||}
\hline
$N_{atoms}$ & $a$ & $a$ & $b$\ (s$^{2}$) & $b$\ (s$^{2}$) & $c$\ (s) & $c$\ (s)\\
\hline
&(formula) & (fit) & (formula) & (fit) & (formula) & (fit)\\
\hline\hline
 10000  & 0.500 & 0.500 &  3475.6 &  3485.2 & 691.7 & 691.7\\
 100000 & 0.500 & 0.500 &  8499.3 &  8672.5 & 691.8 & 691.9\\
 200000 & 0.500 & 0.500 & 11161.0 & 11439.9 & 691.9 & 691.9\\
 500000 & 0.500 & 0.500 & 16015.4 & 16517.2 & 692.1 & 692.1\\
1000000 & 0.500 & 0.500 & 21051.2 & 21821.9 & 692.2 & 692.3\\
2000000 & 0.500 & 0.500 & 27667.0 & 28842.8 & 692.2 & 692.5\\
\hline
 10000  & 0.447 & 0.447 &  3553.9 &  3479.9 & 703.0 & 703.5\\
 100000 & 0.407 & 0.407 &  8718.4 &  8764.5 & 703.2 & 702.6\\
 200000 & 0.385 & 0.385 & 11454.4 & 11584.0 & 703.3 & 702.6\\
 500000 & 0.347 & 0.347 & 16443.3 & 16751.0 & 703.5 & 702.7\\
1000000 & 0.310 & 0.310 & 21618.4 & 22145.7 & 703.6 & 702.8\\
2000000 & 0.268 & 0.268 & 28417.3 & 29283.9 & 703.9 & 702.8\\
\hline\hline
\end{tabular}
\caption{Comparison of the fitted values of the parameters, $a$, $b$, and $c$ with their values predicted by Eqs.\,(\ref{a_formula}), (\ref{b_formula}), and (\ref{c_formula}) when cloud-cloud interactions are turned on.  The top half of the table gives results for a harmonic potential and the bottom half for anharmonic potential.}
\label{table_int_on}
\end{table}

One noticeable feature of these plots is the effect of the anharmonic potential.  The maximum and minimum values of $S_{+}$ vary between approximately 1 and 0 for the harmonic potential while this variation is reduced for the anharmonic potential.  Another smaller effect is that the envelope of the variation of $S_{+}$ decreases more rapidly for the anharmonic potential.

The physical mechanisms for these effects can be identified quantitatively by looking at the data found in Tables\,\ref{table_int_off} and \ref{table_int_on}.  Table\,\ref{table_int_off} compares the fitted values of parameters $a$, $b$, and $c$ with their values determined from Eqs.\,(\ref{a_formula}), (\ref{b_formula}), and (\ref{c_formula}) for six different condensate numbers for the case where cloud-cloud interaction are turn off. The top half of the table contains results for the harmonic potential and the bottom half shows results for the anharmonic potential. Table\,\ref{table_int_on} shows the same comparison when cloud-cloud interactions are on.

The first thing to note is that the approximate formulas for $a$, $b$, and $c$ do a good job in predicting the fitted values for these parameters.  This is the case for all conditions considered.  The formula values for $a$ and $c$ differ from the fitted values by well under 1\% and the largest difference in these values for $b$ is less than 5\%.  Thus these formulas should provide physical insight into the effects of interactions and the anharmonic potential.

One important feature of the comparisons shown in the tables is that the $c$ parameter is insensitive to the number of condensate atoms.  Furthermore, because the fit and formula values for $c$ match well, it shows that the argument, $2\pi c\Omega_{z}$, is the Sagnac phase.  The slightly larger values of $c$ for the anharmonic potential occurs because the anharmonic terms cause the cloud trajectories to move out of the $xy$ plane thus increasing slightly the area of the interferometer over its value for the harmonic potential.  

The $a$ parameter fit and formula comparison is particularly good across all conditions.  The expression for $a$ in Eq.\,(\ref{a_formula}) shows that, in order to maximize $a$ at zero rotation speed, the centers of the two clouds must coincide and their relative velocity $x$ components must be $2v_{B}$ just before the final split. The agreement between fit and formula values for $a$ show that this physical picture for the value of $a$ works for all conditions.

The last parameter to consider is $b$.  We can see from the tables that the value of $b$ is insensitive to the presence of anharmonic terms in the potential. The striking feature of the variation of $b$ for increasing condensate number is that its value increases significantly when cloud-cloud interactions are present over when they are absent.  The main driver of this effect occurs at the second split when two clouds become four clouds.  Without cloud-cloud interactions, the rate of change of the transverse cloud width ($w_{\perp}(t_{1})$) begins to decrease sharply while this rate of change continues to increase when interactions are present. This leads to a significantly larger value of $w_{\perp}(t_{2})$ at the final split.

The $b$ parameter directly measures an effect of the finite-width of the condensate at the time of the final split as shown in Eq.\,(\ref{b_formula}). It is worth noting that its effect on the value of $S_{+}$ is increased for larger rotation speed and, importantly, also for larger interferometer areas.  It is therefore possible that this effect may need to be accounted for in the data analysis of the experimental results for larger-area interferometers.

%%%%%%%%%%%%%%%%%%%%%%%%%%%%%%%%%%%%%%%%%%%%%%%%%%%%%%%%%%%%%
%                  Summary and Conclusions                  %
%%%%%%%%%%%%%%%%%%%%%%%%%%%%%%%%%%%%%%%%%%%%%%%%%%%%%%%%%%%%%
\section{Summary and Conclusions}
\label{summary}

In this work we have applied a variational model providing approximate solutions to the rotating-frame Gross--Pitaevskii equation to a recent dual-Sagnac atom-interferometry measurement of the lab rotation speed. This measurement involved splitting and recombining a small Bose--Einstein condensate in an ideally harmonic potential.  We used the model to study the effects of interactions due to increasing the number of condensate atoms and of the presence of anharmonic terms in the external potential on the operation of the interferometer.

We found that the finite condensate width due to the presence of atom-atom interactions caused a slow decay in the envelope of the variation of the stopped-atom fraction, $S_{\pm}$, with increasing rotating-frame speed, $\Omega_{z}$. The rate of this decay is proportional to the square of the width of the condensate at the time of the final split.  Thus anything that increases this final width will accelerate the envelope decay rate. This envelope decay rate is also proportional to the interferometer area.  Thus we expect that this effect may be important in state-of-the-art interferometer applications.

In our model we found that the final width was affected by the breathing motion of the individual clouds caused by self interactions as well as by interactions between different clouds.  When a condensate cloud is split in two the number of atoms in the daughter clouds is roughly half of the original so that the repulsion is reduced while the confinement from the external potential remains the same.  Thus the cloud will begin to contract increasing the repulsion. Eventually the contraction stops and the cloud begins expanding until the confinement stops the expansion and contraction begins again. Thus the self interaction can lead to a bigger or smaller final width.  We found that interactions between different clouds can moderate the change in the expansion/contraction rate that happens when a cloud is split.  It is possible to minimize the final width by engineering the four-cloud flight time so that the width oscillation is at a minimum.

The amplitude of the envelope of the stopped-atom fraction variation with $\Omega_{z}$ is affected by the presence of anharmonic terms in the potential.  This amplitude (parameter $a$ in the study) is decreased when the two overlapping stopped-atom clouds have a relative velocity and/or are not completely overlapped at the final split.  The relative velocity can be a combination of the relative velocity of the cloud centers and the expansion or contraction of the cloud width. These effects also decrease when the final width decreases.  It is also possible to engineer the trap period to minimize this effect.

Finally, our model predicts that the frequency of the variation of $S_{\pm}$ with $\Omega_{z}$ depends on the Sagnac phase regardless of the presence of interactions and/or anharmonic terms.  We found that $S_{\pm}$ varied sinusoidally with the Sagnac phase under all conditions.  Thus it may be possible to use the fit function in Eq.\,(\ref{fit_function}) to devise a procedure similar to the ellipse-fitting data analysis used in the experiment for common-mode rejection but which can account for interactions and anharmonic effects.

%%%%%%%%%%%%%%%%%%%%%%%%%%%%%%%%%%%%%%%%%%
\vspace{6pt} 
%%%%%%%%%%%%%%%%%%%%%%%%%%%%%%%%%%%%%%%%%%
\authorcontributions{Conceptualization, M.E., C.C. and C.S.; methodology, M.E., C.C. and C.S.; software, M.E.; validation, S.T., C.S., C.H. and A.S.; formal analysis, S.T., C.S., C.H., A.S. and C.C.; investigation, S.T., C.S., C.H., A.S. and C.C.; resources, M.E.; data curation, S.T.; writing--original draft preparation, M.E.; writing--review and editing, M.E., C.C., C.S., S.T., C.S., C.H. and A.S.; visualization, C.H.; supervision, M.E.; project administration, M.E.; funding acquisition, M.E. and C.C.  All authors have read and agreed to the published version of the manuscript.}

%%%%%%%%%%%%%%%%%%%%%%%%%%%%%%%%%%%%%%%%%%
\funding{This research was funded by US National Science Foundation grant number 1707776, DARPA grant number FA9453-19-1-0007 and NASA grant number RSA1549080g.}

%%%%%%%%%%%%%%%%%%%%%%%%%%%%%%%%%%%%%%%%%%
\acknowledgments{The authors would like to thank Jim Stickney for helpful discussions.}

%%%%%%%%%%%%%%%%%%%%%%%%%%%%%%%%%%%%%%%%%%
\conflictsofinterest{The authors declare no conflicts of interest.} 

%%%%%%%%%%%%%%%%%%%%%%%%%%%%%%%%%%%%%%%%%%
%% optional
\abbreviations{The following abbreviations are used in this manuscript:\\

\noindent 
\begin{tabular}{@{}ll}
AI & Atom Interferometry\\
BEC & Bose--Einstein condensate\\
CAL & Cold Atom Laboratory\\
GPE & Gross--Pitaevskii Equation\\
LVM & Lagrangian Variational Method\\
NASA & National Aeronautics and Space Administration\\
RFGPE & Rotation--frame Gross--Pitaevskii equation\\
TOP & Time-averaged orbiting potential
\end{tabular}}

%%%%%%%%%%%%%%%%%%%%%%%%%%%%%%%%%%%%%%%%%%
%% optional
\appendixtitles{yes} 
\appendix

\section{Derivation of the power-law equations of motion}
\label{appA}
We can derive the variational equations of motion  (Eqs.\,(\ref{xj_eom_final})--(\ref{wjeta_eoms_final})) for this power-law external potential by inserting equation (\ref{power_law_potential}) into equation (\ref{partexp}).  Thus we have
\begin{eqnarray*}
U_{\rm ext}^{(3D)}\left({\bf x},{\bf w}\right)
&=&
\sum_{k=1}^{N_{\rm terms}}
C_{p_{x}(k),p_{y}(k),p_{z}(k)}
\sum_{j=1}^{N_{c}}
\left(
\frac{1}{\pi^{3/2}w_{jx}w_{jy}w_{jz}}
\right)
I_{p_{x}(k)}(x_{j},w_{jx})
I_{p_{y}(k)}(y_{j},w_{jy})
I_{p_{z}(k)}(z_{j},w_{jz})
\end{eqnarray*}
where we have defined
\begin{equation}
I_{k}(\eta_{0},w_{0}) \equiv 
\int_{-\infty}^{\infty}\,
\eta^{k} e^{-\left(\eta-\eta_{0}\right)^{2}/w_{0}^{2}}\,d\eta
\equiv
\left(w_{0}\pi^{1/2}\right)
J_{k}\left(\eta_{0},w_{0}\right),
\quad
k = 0,1,2,\dots
\end{equation}
This integral is easily evaluated so that e can write the variational potential in a compact form using the newly defined function $J_{k}(\eta,w)$.  The result is
\begin{eqnarray}
U_{\rm ext}^{(3D)}\left({\bf x},{\bf w}\right)
&=& 
\sum_{j=1}^{N_{c}}
\sum_{k=1}^{N_{\rm terms}}
C_{p_{x}(k),p_{y}(k),p_{z}(k)}
J_{p_{x}(k)}(x_{j},w_{jx})
J_{p_{y}(k)}(y_{j},w_{jy})
J_{p_{z}(k)}(z_{j},w_{jz})
\label{uextsc}
\end{eqnarray}
where $J_{k}(\eta,w)$ can be written as
\begin{equation}
J_{k}(\eta,w) = 
\vast\{
\begin{array}{lc}
\sum_{m=0}^{k/2}
\left(
\frac
{k!}
{m!(k-2m)!}
\right)\eta^{k-2m}
(\tfrac{1}{2}w)^{2m} & {\rm k=even\ integer}\\
\sum_{m=0}^{(k-1)/2}
\left(
\frac
{k!}
{m!(k-2m)!}
\right)\eta^{k-2m}
(\tfrac{1}{2}w)^{2m} & {\rm k=odd\ integer}
\end{array}
\label{var_plp}
\end{equation}
We can use this expression to write down the equations of motion.

For the equations of motion we need to compute the derivative of $U_{\rm ext}({\bf x},{\bf w})$ with respect to the center coordinates and center widths. These are easily done with results
\begin{eqnarray*}
\frac{\partial U_{\rm ext}^{(3D)}}{\partial x_{j}}
&=&
\sum_{k=1}^{N_{\rm terms}}
C_{p_{x}(k),p_{y}(k),p_{z}(k)}
\left(\frac{\partial J_{p_{x}(k)}}{\partial x_{j}}\right)
J_{p_{y}(k)}(y_{j},w_{jy})
J_{p_{z}(k)}(z_{j},w_{jz})\\
\frac{\partial U_{\rm ext}^{(3D)}}{\partial y_{j}}
&=&
\sum_{k=1}^{N_{\rm terms}}
C_{p_{x}(k),p_{y}(k),p_{z}(k)}
J_{p_{x}(k)}(x_{j},w_{jx})
\left(\frac{\partial J_{p_{y}(k)}}{\partial y_{j}}\right)
J_{p_{z}(k)}(z_{j},w_{jz})\\
\frac{\partial U_{\rm ext}^{(3D)}}{\partial z_{j}}
&=&
\sum_{k=1}^{N_{\rm terms}}
C_{p_{x}(k),p_{y}(k),p_{z}(k)}
J_{p_{x}(k)}(x_{j},w_{jx})
J_{p_{y}(k)}(y_{j},w_{jy})
\left(\frac{\partial J_{p_{z}(k)}}
{\partial z_{j}}\right)
\end{eqnarray*}
\begin{eqnarray*}
\frac{\partial U_{\rm ext}^{(3D)}}{\partial w_{jx}}
&=&
\sum_{k=1}^{N_{\rm terms}}
C_{p_{x}(k),p_{y}(k),p_{z}(k)}
\left(\frac{\partial J_{p_{x}(k)}}{\partial w_{jx}}\right)
J_{p_{y}(k)}(y_{j},w_{jy})
J_{p_{z}(k)}(z_{j},w_{jz})\\
\frac{\partial U_{\rm ext}^{(3D)}}{\partial w_{jy}}
&=&
\sum_{k=1}^{N_{\rm terms}}
C_{p_{x}(k),p_{y}(k),p_{z}(k)}
J_{p_{x}(k)}(x_{j},w_{jx})
\left(\frac{\partial J_{p_{y}(k)}}{\partial w_{jy}}\right)
J_{p_{z}(k)}(z_{j},w_{jz})\\
\frac{\partial U_{\rm ext}^{(3D)}}{\partial w_{jz}}
&=&
\sum_{k=1}^{N_{\rm terms}}
C_{p_{x}(k),p_{y}(k),p_{z}(k)}
J_{p_{x}(k)}(x_{j},w_{jx})
J_{p_{y}(k)}(y_{j},w_{jy})
\left(\frac{\partial J_{p_{z}(k)}}
{\partial w_{jz}}\right).
\end{eqnarray*}
where the derivatives can be expressed as follows.
\begin{equation}
\frac{\partial J_{k}}{\partial \eta} = 
\vast\{
\begin{array}{lc}
\sum_{m=0}^{k/2}
\left(
\frac
{k!}
{m!(k-2m-1)!}
\right)\eta^{k-2m-1}
(\tfrac{1}{2}w)^{2m} & {\rm k=even\ integer}\\
\sum_{m=0}^{(k-1)/2}
\left(
\frac
{k!}
{m!(k-2m-1)!}
\right)\eta^{k-2m-1}
(\tfrac{1}{2}w)^{2m} & {\rm k=odd\ integer}
\end{array}
\end{equation}
and where $\eta=x,y,z$ and
\begin{equation}
\frac{\partial J_{k}}{\partial w} = 
\vast\{
\begin{array}{lc}
\sum_{m=1}^{k/2}
\left(
\frac
{k!}
{(m-1)!(k-2m)!}
\right)\eta^{k-2m}
(\tfrac{1}{2}w)^{2m-1} & {\rm k=even\ integer}\\
\sum_{m=1}^{(k-1)/2}
\left(
\frac
{k!}
{(m-1)!(k-2m)!}
\right)\eta^{k-2m}
(\tfrac{1}{2}w)^{2m-1} & {\rm k=odd\ integer}
\end{array}
\end{equation}

\section{Interaction terms in the variational equations of motion}
\label{appB}

The space and width gradients of $U_{\rm int}^{(3D)}({\bf x},{\bf w})$ appearing in the variational equations of motion are derived in Ref.\,\cite{PhysRevA.99.043615}. The result for the space gradients is 
\begin{eqnarray}
F_{j\eta}({\bf x},{\bf w}) \equiv 
\frac{\partial U_{\rm int}^{(3D)}}{\partial \eta_{j}}
&=&
\left(
\frac{4gN}{(\pi)^{3/2}N_{c}}
\right)
\sum_{\substack{j^{\prime}=1\\j^{\prime}\ne j}}^{N_{c}}
\prod_{\eta^{\prime}=x,y,z}
\left(
\frac
{
\exp
\left\{
-\frac
{\left(\eta^{\prime}_{j^{\prime}}-\eta^{\prime}_{j}\right)^{2}}
{w_{j^{\prime}\eta^{\prime}}^{2}+w_{j\eta^{\prime}}^{2}}
\right\}
}
{
\left(w_{j^{\prime}\eta^{\prime}}^{2}+w_{j\eta^{\prime}}^{2}\right)^{1/2}
}
\right)
\left(
\frac
{\eta_{j^{\prime}}-\eta_{j}}
{w_{j^{\prime}\eta}^{2}+w_{j\eta}^{2}}
\right),\nonumber\\
{\rm\ where\ }\eta 
&=& x,y,z
\label{U_int_space_final}
\end{eqnarray}
It is worth noting that this quantity describes the ``force'' exerted on cloud $j$ due to all of the other clouds $j^{\prime}\ne j$ and is only non-zero when there is significant overlap between two different clouds. It is also notable that the force of cloud $j^{\prime}$ acting on cloud $j$ is equal and opposite to the force of cloud $j$ on $j^{\prime}$ as can be seen from the above equation.  Thus all of the cloud-cloud interactions obey a ``Newton's Third Law'' condition.

The equations of motion contain terms which are twice the width gradient of $U_{\rm int}^{(3D)}({\bf x},{\bf w})$. These can be written in the following form:
\begin{eqnarray}
2\frac{\partial U_{\rm int}}{\partial w_{j\eta}}
&\equiv&
-\frac{gN}{(2\pi)^{3/2}N_{c}}
\left(
\frac{1}{w_{jx}w_{jy}w_{jz}w_{j\eta}}
\right) +
W_{j\eta}({\bf x},{\bf w}),
\quad
\eta = x,y,z. 
\label{U_int_width_deriv_final}
\end{eqnarray}
where the first term accounts for the self-interaction of cloud $j$. The second term accounts for the interaction of cloud $j$ with the other clouds and has the form\,\cite{PhysRevA.99.043615}
\begin{equation}
W_{j\eta}({\bf x},{\bf w}) \equiv
\frac{4gN}{(\pi)^{3/2}N_{c}}
\sum_{\substack{j^{\prime}=1\\j^{\prime}\ne j}}^{N_{c}}
\prod_{\eta^{\prime}=x,y,z}
\left(
\frac
{
\exp
\left\{
-\frac
{\left(\eta^{\prime}_{j^{\prime}}-\eta^{\prime}_{j}\right)^{2}}
{w_{j^{\prime}\eta^{\prime}}^{2}+w_{j\eta^{\prime}}^{2}}
\right\}
}
{
\left(w_{j^{\prime}\eta^{\prime}}^{2}+w_{j\eta^{\prime}}^{2}\right)^{1/2}
}
\right)
\left(
\frac
{
w_{j\eta}
\left(
2\left(\eta_{j^{\prime}}-\eta_{j}\right)^{2} -
\left(w_{j^{\prime}\eta}^{2}+w_{j\eta}^{2}\right)
\right)
}
{
\left(w_{j^{\prime}\eta}^{2}+w_{j\eta}^{2}\right)^{2}
}
\right).
\label{width_int}
\end{equation}
Note that this term is negligible unless cloud $j$ overlaps one or more of the other clouds, $j^{\prime}\ne j$.

%%%%%%%%%%%%%%%%%%%%%%%%%%%%%%%%%%%%%%%%%%
\reftitle{References}

\externalbibliography{yes}
\bibliography{mdpi.bib}

\end{document}